\documentclass[twocolumn,showpacs]{revtex4}
\usepackage{graphicx}
\usepackage{amsmath}
\usepackage{bm}
\begin{document}

\title{Splitting of a Cooper pair by a pair of Majorana bound states}
\author{Johan Nilsson}
\affiliation{Instituut-Lorentz, Universiteit Leiden, P.O. Box 9506, 2300 RA Leiden, The Netherlands}
\author{A. R. Akhmerov}
\affiliation{Instituut-Lorentz, Universiteit Leiden, P.O. Box 9506, 2300 RA Leiden, The Netherlands}
\author{C. W. J. Beenakker}
\affiliation{Instituut-Lorentz, Universiteit Leiden, P.O. Box 9506, 2300 RA Leiden, The Netherlands}
\date{August 2008}
\begin{abstract}
Majorana bound states are spatially localized superpositions of electron and hole excitations in the middle of a superconducting energy gap. A single qubit can be encoded nonlocally in a pair of spatially separated Majorana bound states. Such Majorana qubits are in demand as building blocks of a topological quantum computer, but direct experimental tests of the nonlocality remain elusive. Here we propose a method to probe the nonlocality by means of crossed Andreev reflection, which is the injection of an electron into one bound state followed by the emission of a hole by the other bound state (equivalent to the splitting of a Cooper pair over the two states). We have found that, at sufficiently low excitation energies, this nonlocal scattering process dominates over local Andreev reflection involving a single bound state. As a consequence, the low-temperature and low-frequency fluctuations $\delta I_{i}$ of currents into the two bound states $i=1,2$ are maximally correlated: $\overline{\delta I_{1}\delta I_{2}}=\overline{\delta I_{i}^{2}}$.
\end{abstract}
\maketitle

\begin{figure}[tb]
\centerline{\includegraphics[width=0.9\linewidth]{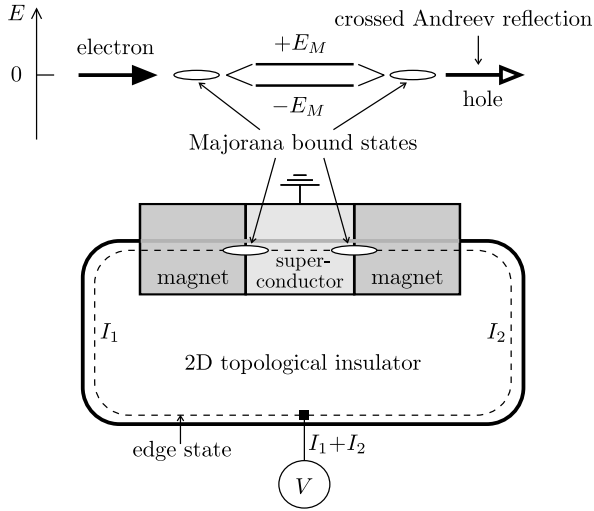}}
\caption{\label{MIM_layout}
Top panel: Energy diagram of two Majorana bound states (levels at zero energy), which split into a pair of levels at $\pm E_{M}$ upon coupling. Whether the upper level is excited or not determines the states $|1\rangle$ and $|0\rangle$ of a qubit. Crossed Andreev reflection probes the nonlocality of this Majorana qubit. Lower panel: Detection of crossed Andreev reflection by correlating the currents $I_{1}$ and $I_{2}$ that flow into a superconductor via two Majorana bound states.
}
\end{figure}

Majorana bound states are coherent superpositions of electron and hole excitations of zero energy, trapped in the middle of the superconducting energy gap by a nonuniformity in the pair potential. Two Majorana bound states nonlocally encode a single qubit (see Fig.\ \ref{MIM_layout}, top panel). If the bound states are widely separated, the qubit is robust against local sources of decoherence and provides a building block for topological quantum computation \cite{Kit00,Nay07}.

While Majorana bound states have not yet been demonstrated experimentally, there is now a variety of candidate systems. In an $s$-wave superconductor, zero-point motion prevents the formation of bound states at zero energy. Early proposals for Majorana bound states therefore considered $p$-wave superconductors \cite{Rea00,Iva01}, with ${\rm Sr}_{2}{\rm RuO}_{4}$ as a candidate material \cite{Sar06}, or $p$-wave superfluids formed by fermionic cold atoms \cite{Tew07}. More recently, it was discovered \cite{Fu08a,Gha07,Ber08} that Majorana bound states can be induced by $s$-wave superconductivity in a metal with a Dirac spectrum (such as graphene or the boundary of a topological insulator). Several tunneling experiments have been proposed \cite{Sem07,Bol07,Tew08} to search for the Majorana bound states predicted to occur in these systems.

Here we show that crossed Andreev reflection \cite{Bye95,Har96,Mar96} by a pair of Majorana bound states is a direct probe of the nonlocality. \textit{Crossed} Andreev reflection is the nonlocal conversion of an electron excitation into a hole excitation, each in a separate lead. \textit{Local} Andreev reflection, in contrast, converts an electron into a hole in the same lead. Equivalently, local Andreev reflection injects a Cooper pair in a single lead, while crossed Andreev reflection splits a Cooper pair over two leads. We have found that at sufficiently low excitation energies, \textit{local Andreev reflection by a pair of Majorana bound states is fully suppressed in favor of crossed Andreev reflection.}

The suppression is not a property of the dispersion relation in the leads (as in Refs.\ \cite{Cay08,Ben08}), but directly probes the Majorana character of the Hamiltonian  \cite{Nay07}
\begin{equation}
H_{M}=iE_{M}\gamma_{1}\gamma_{2}, \label{HMdef}
\end{equation}
of the pair of weakly coupled bound states (labeled $1$ and $2$). The $\gamma_{i}$'s are Majorana operators, defined by $\gamma_{i}=\gamma_{i}^{\dagger}$, $\gamma_{i}\gamma_{j}+\gamma_{j}\gamma_{i}=2\delta_{ij}$.  The coupling energy $E_{M}$ splits the two zero-energy levels into a doublet at $\pm E_{M}$. The suppression of local Andreev reflection happens when the width $\Gamma_{M}$ of the levels in the doublet (which is finite because of leakage into the leads) and the excitation energy $E$ are both $\ll E_{M}$. (The relative magnitude of $\Gamma_{M}$ and $E$ does not matter.)

Our theoretical analysis is particularly timely in view of recent advances in the experimental realization of topological insulators in two-dimensional (2D) HgTe quantum wells \cite{Koe07} and 3D BiSb crystals \cite{Hsi08}. Topological insulators are characterized by an inverted band gap, which produces metallic states at the interface with vacuum or any material with a normal (noninverted) band gap \cite{topins}. The metallic states are 2D surface states if the insulator is 3D, while if the insulator is 2D the metallic states are 1D edge states. 

These recent experiments \cite{Koe07,Hsi08} used nonsuperconducting electrodes. A superconducting proximity effect between Nb and BiSb was reported in earlier work \cite{Kas96}, so that we expect a search for the predicted \cite{Fu08a} Majorana bound states to be carried out in the near future. Anticipating these developments, we will identify observable consequences of the suppression of local Andreev reflection, by calculating the shot noise in a 2D topological insulator with a superconducting electrode (Fig.\ \ref{MIM_layout}, lower panel). A similar calculation can be done for the 3D case, and indeed our conclusions are quite general --- as we will now demonstrate by showing that the Majorana Hamiltonian \eqref{HMdef} directly implies the suppression of local Andreev reflection.

For this purpose write the unitary scattering matrix $S(E)$ in a model-independent form,
\begin{equation}
S(E)=1+2\pi i W^{\dagger}(H_{M}-E-i\pi WW^{\dagger})^{-1}W, \label{SHMrelation}
\end{equation}
with $W$ the matrix that describes the coupling of the scatterer (Hamiltonian $H_{M}$) to the leads. In our case, we have
\begin{equation}
W=\begin{pmatrix}
w_{1}&0&w_{1}^{\ast}&0\\
0&w_{2}&0&w_{2}^{\ast}
\end{pmatrix},\;\;
H_{M}=\begin{pmatrix}
0&iE_{M}\\
-iE_{M}& 0
\end{pmatrix}.\label{WHMdef}
\end{equation}
The expression for $H_{M}$ is Eq.\ \eqref{HMdef} in the basis $\{\Phi_{1},\Phi_{2}\}$ of the two Majorana bound states, while $W$ is the coupling matrix in the basis $\{\Phi_{e,1},\Phi_{e,2},\Phi_{h,1},\Phi_{h,2}\}$ of propagating electron and hole modes in leads $1$ and $2$. We have assumed that lead $1$ is coupled only to bound state $1$ and lead $2$ only to bound state $2$, and we have also assumed that the energy dependence of the coupling amplitudes $w_{i}$ can be neglected. (In the exact calculation given later on for a specific model, neither assumption will be made.) Without loss of generality we can choose the $w_{i}$'s to be purely real numbers by adjusting the phases of the basis states in the leads.

Substitution of Eq.\ \eqref{WHMdef} into Eq.\ \eqref{SHMrelation} gives the electron and hole blocks of the scattering matrix,
\begin{equation}
S\equiv\begin{pmatrix}
s^{ee}&s^{eh}\\
s^{he}&s^{hh}
\end{pmatrix}=
\begin{pmatrix}
1+A&A\\
A&1+A
\end{pmatrix},\label{Sblocks}
\end{equation}
which turn out to depend on a single $2\times 2$ matrix $A$ with elements
\begin{equation}
A=Z^{-1}\begin{pmatrix}
i \Gamma_{1}(E+i\Gamma_{2})&
-E_{M}\sqrt{\Gamma_{1}\Gamma_{2}}\\
E_{M}\sqrt{\Gamma_{1}\Gamma_{2}}&
i \Gamma_{2}(E+i\Gamma_{1})
\end{pmatrix}.\label{Adef}
\end{equation}
We have abbreviated
\begin{equation}
Z=E_{M}^{2}-(E+i\Gamma_{1})(E+i\Gamma_{2}),\;\;
\Gamma_{i}=2\pi w_{i}^{2}.\label{ZGammadef}
\end{equation}
(The width $\Gamma_{M}$ introduced earlier equals $\Gamma_{1}+\Gamma_{2}$.) Unitarity of $S$ is guaranteed by the identity
\begin{equation}
A+A^{\dagger}+2AA^{\dagger}=0.\label{Aidentity}
\end{equation}

In the limit of low excitation energies and weak coupling to the leads, this simplifies to
\begin{equation}
A\approx\frac{\sqrt{\Gamma_{1}\Gamma_{2}}}{E_{M}}\begin{pmatrix}
0&-1\\
1&0
\end{pmatrix},\;\;\mbox{for}\;\;E,\Gamma_{i}\ll E_{M}.\label{Asimple}
\end{equation}
The scattering matrix $s^{he}=A$ that describes Andreev reflection of an electron into a hole has therefore only off-diagonal elements in this limit, so only \textit{crossed} Andreev reflection remains. More specifically, an electron incident in lead $1$ is transferred to the other lead $2$ either as an electron or as a hole, with equal probabilities $p=\Gamma_{1}\Gamma_{2}/E_{M}^{2}$. The probability for local Andreev reflection is smaller than the probability $p$ for crossed Andreev reflection by a factor $(\Gamma_{1}/\Gamma_{2})(E^{2}/E_{M}^{2}+\Gamma_{2}^{2}/E_{M}^{2}) \ll 1$.

Because the probabilities to transfer to the other lead as an electron or as a hole are the same, crossed Andreev reflection cannot be detected in the time averaged current $\bar{I}_{i}$ in lead $i$, but requires measurement of the current fluctuations $\delta I_{i}(t)=I_{i}(t)-\bar{I}_{i}$. We consider the case that both leads are biased equally at voltage $V$, while the superconductor is grounded. At low temperatures $T\ll eV/k_{B}$ the current fluctuations are dominated by shot noise. In the regime $p\ll 1$ of interest, this noise consists of independent current pulses with Poisson statistics \cite{Bla00}. The Fano factor (ratio of noise power and mean current) measures the charge transferred in a current pulse.

The total (zero frequency) noise power $P=\sum_{ij}P_{ij}$, with
\begin{equation}
P_{ij}=\int_{-\infty}^{\infty}dt\,\overline{\delta I_{i}(0)\delta I_{j}(t)},\label{Pijdef}
\end{equation}
has Fano factor $F=P/e\bar{I}$ (with $\bar{I}=\sum_{i}\bar{I}_{i}$) equal to 2 rather than equal to 1 because the superconductor can only absorb electrons in pairs \cite{Jon94}. As we will now show, the suppression of local Andreev reflection by the pair of Majorana bound states produces a characteristic signature in the individual noise correlators $P_{ij}$.

The general expressions for $\bar{I}_{i}$ and $P_{ij}$ in terms of the scattering matrix elements are \cite{Ana96}
\begin{align}
\bar{I}_{i}={}&\frac{e}{h}\int_{0}^{eV}dE\,\bigl(1-{\cal R}^{ee}_{ii}+{\cal R}^{hh}_{ii}\bigr),\label{barIi}\\
P_{ij}={}&\frac{e^{2}}{h}\int_{0}^{eV}dE\,{\cal P}_{ij}(E),\label{Pij}
\end{align}
with the definitions
\begin{align}
{\cal P}_{ij}(E)={}&\delta_{ij}{\cal R}^{ee}_{ii}+\delta_{ij}{\cal R}^{hh}_{ii}-{\cal R}^{ee}_{ij}{\cal R}^{ee}_{ji}-{\cal R}^{hh}_{ij}{\cal R}^{hh}_{ji}\nonumber\\
&+{\cal R}^{eh}_{ij}{\cal R}^{he}_{ji}+{\cal R}^{he}_{ij}{\cal R}^{eh}_{ji},\label{calPij}\\
{\cal R}^{xy}_{ij}(E)={}&\sum_{k}s_{ik}^{xe}(E)[s_{jk}^{ye}(E)]^{\ast},\;\;x,y\in\{e,h\}.\label{Rxy}
\end{align}
Substitution of the special form \eqref{Sblocks} of $S$ for the pair of Majorana bound states, results in
\begin{align}
\bar{I}_{i}={}&\frac{2e}{h}\int_{0}^{eV}dE\,(AA^{\dagger})_{ii},\label{barIi2}\\
P_{ij}={}&e\bar{I}_{i}\delta_{ij}+\frac{2e^{2}}{h}\int_{0}^{eV}dE\,\bigl[\,
|A_{ij}+(AA^{\dagger})_{ij}|^{2}\nonumber\\
&\qquad\qquad\qquad\qquad-|(AA^{\dagger})_{ij}|^{2}\,\bigr],\label{Pij2}
\end{align}
where we have used the identity \eqref{Aidentity}.

We now take the low energy and weak coupling limit, where $A$ becomes the off-diagonal matrix \eqref{Asimple}. Then we obtain the remarkably simple result
\begin{equation}
P_{ij}=e\bar{I}_{1}=e\bar{I}_{2}=\frac{e\bar{I}}{2},\;\;\mbox{for}\;\;eV,\Gamma_{i}\ll E_{M}.\label{PIresult}
\end{equation}
The total noise power $P\equiv\sum_{ij}P_{ij}=2e\bar{I}$ has Fano factor two, as it should be for transfer of Cooper pairs into a superconductor \cite{Jon94}, but the noise power of the separate leads has unit Fano factor: $F_{i}\equiv P_{ii}/e\bar{I}_{i}=1$. Because local Andreev reflection is suppressed, the current pulses in a single lead transfer charge $e$ rather than $2e$ into the superconductor. The positive cross-correlation of the current pulses in the two leads ensures that the total transferred charge is $2e$. This ``splitting'' of a Cooper pair is a highly characteristic signature of a Majorana qubit, reminiscent of the $h/e$ (instead of $h/2e$) flux periodicity of the Josephson effect \cite{Kit00,Kwo04,Fu08b}.

Notice that for any stochastic process the cross-correlator is bounded by the auto-correlator,
\begin{equation}
|P_{12}| \leq \tfrac{1}{2}(P_{11}+P_{22}).\label{Pinequality}
\end{equation}
The positive cross-correlation \eqref{PIresult} is therefore maximally large. This is a special property of the low energy, weak coupling limit. There is no inconsistency with the conclusion of Bolech and Demler \cite{Bol07}, that the currents into two Majorana bound states fluctuate independently, because that conclusion applies to the regime $eV\gg E_{M}$. The duration $\hbar/eV$ of the current pulses is then shorter than the time $\hbar/E_{M}$ needed to transfer charge between the bound states, so no cross-correlations can develop. In this high-voltage regime the two Majorana bound states behave as independent Andreev resonances, for which the noise correlators are known \cite{Fau98},
\begin{equation}
P_{ii}=e\bar{I}_{i},\;\;P_{12}=0,\;\;\mbox{for}\;\;eV\gg E_{M},\Gamma_{i}.\label{PIresult0}
\end{equation}
While the Fano factors of the individual leads $F_{i}=1$ remain the same, the total noise power $P\equiv\sum_{ij}P_{ij}=e\bar{I}$ has Fano factor $F=1$ rather than $F=2$ when the cross-correlator $P_{12}$ vanishes in the high-voltage regime.

As a specific model that can be solved exactly and is experimentally relevant, we consider a 2D topological insulator contacted at the edge by one superconducting electrode in between a pair of magnets (Fig.\ \ref{MIM_layout}, bottom panel). As discoverd by Fu and Kane \cite{Fu08a}, a Majorana bound state appears at the intersection of the magnet--superconductor interface with the edge of the insulator. The four-component wave function $\Psi=(\Psi_{e\uparrow},\Psi_{e\downarrow},\Psi_{h\uparrow},\Psi_{h\downarrow})$ of the edge state satisfies \cite{Fu08a}
\begin{equation}
\begin{pmatrix}
\bm{m}\cdot\bm{\sigma}+vp\sigma_{z}-E_{F}&\Delta\\
\Delta^{\ast}&\bm{m}\cdot\bm{\sigma}-vp\sigma_{z}+E_{F}
\end{pmatrix}\Psi=E\Psi.\label{HDBdG}
\end{equation}
Here $p=-i\hbar\partial/\partial x$ is the momentum operator, $E_{F}$ the Fermi energy, $v$ the Fermi velocity, $\Delta$ the superconducting pair potential, $\bm{m}$ the magnetization vector, and $\bm{\sigma}=(\sigma_{x},\sigma_{y},\sigma_{z})$ the vector of Pauli matrices (acting in the space of right and left movers $\uparrow,\downarrow$).

We set $\Delta(x)=0$ everywhere except $\Delta=\Delta_{0}$ for $0<x<l_{0}$. We also set $\bm{m}(x)=0$ everywhere except $\bm{m}=(m_{0},0,0)$ for $-l_{1}<x<0$ and $\bm{m}=(m_{0}\cos\phi,m_{0}\sin\phi,0)$ for $l_{0}<x<l_{0}+l_{2}$. We assume that $|m_{0}|>|E_{F}|$, so that the Fermi level lies in a gap in the magnets as well as in the superconductor. The decay length in the superconductor is the coherence length $\xi_{0}=\hbar v/\Delta_{0}$, while the decay length in the magnets is given by $\lambda_{0}=\hbar v(m_{0}^{2}-E_{F}^{2})^{-1/2}$. For $\lambda_{0}\lesssim\xi_{0}$ the only bound state at the magnet--superconductor interface is the zero-energy Majorana state.

\begin{figure}[tb]
\centerline{\includegraphics[width=0.8\linewidth]{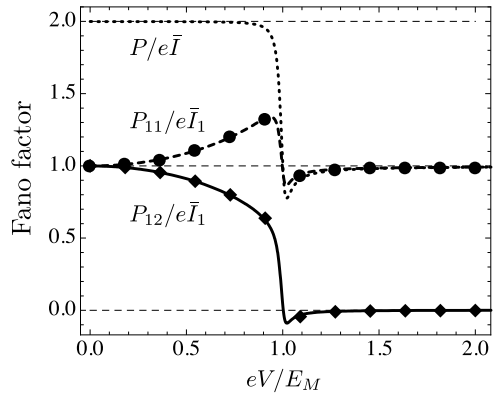}}
\caption{\label{fig_results}
Data points: Auto-correlator $P_{11}$ (circles) and cross-correlator $P_{12}$ (diamonds) of the current fluctuations for the model Hamiltonian \eqref{HDBdG}. The parameters chosen are $E_{F}=0$, $\phi=0$, $m_{0}/\Delta_{0}=1$, $l_{0}=2.3\,\xi_{0}$, $l_{1}=l_{2}=3\,\xi_{0}$. The correlators are normalized by $e\bar{I}_{1}$, to demonstrate the low- and high-voltage limits \eqref{PIresult} and \eqref{PIresult0}. The dashed and solid curves result from the model-independent scattering matrix \eqref{SHMrelation}, with the parameters given by Eqs.\ \eqref{EMresult} and \eqref{Gammaresult}. The dotted curve is the corresponding result for the total noise power $P=\sum_{ij}P_{ij}$, normalized by $e\bar{I}=e\sum_{i}\bar{I}_{i}$. 
}
\end{figure}

We have calculated the scattering states for this model by matching the $\Psi$'s at the opposite sides of the four interfaces $x=-l_{1},0,l_{0},l_{0}+l_{2}$. The resulting scattering matrix is then substituted in the general expressions (\ref{barIi}--\ref{Rxy}) to obtain the zero-temperature, zero-frequency noise correlators as a function of the applied voltage $V$. Representative results are shown in Fig.\ \ref{fig_results} (data points). At low voltages we confirm the unit Fano factor and maximal cross-correlation of Eq.\ \eqref{PIresult}, obtained from the model-independent scattering matrix \eqref{SHMrelation}. Also the crossover to the conventional high-voltage regime \eqref{PIresult0} of independent resonances is clearly visible.

For a quantitative comparison of the two calculations we need the splitting and broadening of the Majorana bound states in the tunneling regime $l_{1},l_{2}\gg\lambda_{0}$, $l_{0}\gg\xi_{0}$. We find
\begin{align}
&E_{M}=e^{-l_{0}/\xi_{0}}
\cos \Bigl[ \frac{\phi}{2} + \frac{E_{F} l_{0}}{\hbar v} 
+ \arctan \bigl( \frac{E_{F} \lambda_{0}}{\hbar v} \bigr) \Bigl]
\frac{2\hbar v}{\xi_{0}+\lambda_{0}}, \label{EMresult}\\
&\Gamma_{i}=e^{-2l_{i}/\lambda_{0}}(1-E_{F}^{2}/m_{0}^{2})\frac{2\hbar v}{\xi_{0}+\lambda_{0}}.\label{Gammaresult}
\end{align}
Notice that the level splitting can be controlled by varying the angle $\phi$ between the magnetizations at the two sides of the superconductor \cite{note1}. In Fig.\ \ref{fig_results} we use these parameters to compare the model-independent calculation based on the scattering matrix \eqref{SHMrelation} (curves) with the results from the model Hamiltonian \eqref{HDBdG} (data points), and find excellent agreement.

The setup sketched in Fig.~\ref{MIM_layout} might be realized in a HgTe quantum well \cite{Koe07}. 
The relevant parameters for this material are as follows. The gap in the bulk insulator is of the order of 20~meV and the magnetic gap can be as large as 3~meV at a magnetic field of 1~T. The smallest energy scale is therefore the gap induced by the superconductor, estimated \cite{Fu08b} at $\Delta_{0}=0.1\,{\rm meV}$. With $\hbar v=0.36\,{\rm meV}\cdot\mu{\rm m}$ this gives a superconducting coherence length of $\xi_{0}=3.6\,\mu{\rm m}$, comparable to the magnetic penetration length $\lambda_{0}$ at a field of 0.03~T. For the calculation in Fig.\ \ref{fig_results} we took $\xi_{0}=\lambda_{0}$ and then took the length $l_{0}$ of the superconducting contact equal to $2.3\,\xi_{0}\simeq 8\,\mu{\rm m}$,
and the lengths $l_{1},l_{2}$ of the magnets both equal to $3\,\xi_{0}\simeq 11\,\mu{\rm m}$. The level splitting is then $E_{M}=0.1\,\Delta_{0}=10\,\mu{\rm eV}\cong 100\,{\rm mK}$. At a temperature of the order of 10~mK we would then have a sufficiently broad range of voltages where $k_{B}T<eV<E_{M}$.

In conclusion, we have demonstrated the suppression of local Andreev reflection by a pair of Majorana bound states at low excitation energies. The remaining crossed Andreev reflection amounts to the splitting of a Cooper pair over the two spatially separated halves of the Majorana qubit. This nonlocal scattering process has a characteristic signature in the maximal positive cross-correlation ($P_{12}=P_{11}=P_{22}$) of the current fluctuations. The splitting of a Cooper pair by the Majorana qubit produces a pair of excitations in the two leads that are maximally entangled in the momentum (rather than the spin) degree of freedom, and might be used as ``flying qubits'' in quantum information processing.

We acknowledge discussions with F. D. M. Haldane, J. E. Moore, and P. Recher. This research was supported by the Dutch Science Foundation NWO/FOM.


\begin{thebibliography}{99}
\bibitem{note1} With respect to the level splitting, the angle $\phi$ between the magnetizations plays the same role as the superconducting phase difference in the Josephson junction of Ref.\ \protect\cite{Fu08b}. One can indeed derive an exact duality relation for the Hamiltonian \eqref{HDBdG} under the interchange $(m_{x},m_{y},m_{z})\leftrightarrow({\rm Re}\,\Delta,{\rm Im}\,\Delta,E_{F})$.
\end{thebibliography}
\end{document}